\documentclass[final,times,sort&compress,onecolumn]{elsarticle}
\usepackage[margin=1in]{geometry}
\usepackage[utf8]{inputenc}
\usepackage{bm}
\usepackage[dvipsnames]{xcolor}
\usepackage{multicol}
\usepackage{multirow}
\usepackage[normalem]{ulem}
\usepackage{subfig}
\usepackage{graphicx}
\usepackage{amsmath}
\usepackage{slashed}
\usepackage{hyperref}
\allowdisplaybreaks

\def\slash#1{\setbox0=\hbox{$#1$}               
        \dimen0=\wd0                            
        \setbox1=\hbox{/} \dimen1=\wd1          
        \ifdim\dimen0>\dimen1                   
        \rlap{\hbox to \dimen0{\hfil/\hfil}}    
        #1                                      
        \else              
        \rlap{\hbox to \dimen1{\hfil$#1$\hfil}} 
        /                                       
        \fi}                                    %

\renewcommand*{\thefootnote}{\fnsymbol{footnote}}

\begin{document}

\rightline{JLAB-THY-25-4300}

\title{Transverse single-spin asymmetries in $\gamma$SIDIS as a direct probe of quark-gluon-quark longitudinal momentum structure}

\author[label1]{Michael Harris} \ead{mah017@lvc.edu}
\author[label1]{Jacob Marsh} \ead{jam022@lvc.edu}
\author[label1]{Daniel Pitonyak} \ead{pitonyak@lvc.edu}
\author[label2,label3]{Alexei Prokudin} \ead{prokudin@jlab.org}
\author[label1]{Jack Putnam} \ead{jdp002@lvc.edu}
\author[label4]{Daniel Rein} \ead{Daniel\_M.Rein@gmx.de}
\author[label4]{Marc Schlegel} \ead{marc.schlegel@uni-tuebingen.de}

\address[label1]{Department of Physics, Lebanon Valley College, Annville, Pennsylvania 17003, USA}
\address[label2]{Division of Science, Penn State University Berks, Reading, Pennsylvania 19610, USA}
\address[label3]{Jefferson Lab, Newport News, Virginia 23606, USA}
\address[label4]{Institute for Theoretical Physics, University of T\"{u}bingen, Auf der Morgenstelle 14, D-72076, T\"{u}bingen, Germany}

\begin{abstract}
\noindent  Transverse single-spin asymmetries in the semi-inclusive deep-inelastic production of isolated photons ($\gamma$SIDIS), $A_{UT}^{\gamma {\rm SIDIS}}$,  provide an unprecedented opportunity to extract the quark-gluon-quark correlators $F_{FT}(x,x')$ and $G_{FT}(x,x')$ point-by-point in their full support $x,x'$.  We utilize realistic models for these functions, based on input from the Sivers transverse momentum dependent parton distribution function and imposing constraints from the $d_2$ matrix element calculated in lattice QCD, in order to provide numerical estimates for $A_{UT}^{\gamma {\rm SIDIS}}$ at the Electron-Ion Collider (EIC).  We thoroughly explore the EIC phase space in order to isolate in which regions the asymmetry can be sizable, finding it can be as much as $10\%$ or larger for certain kinematics.  Given that $F_{FT}(x,x')$ and $G_{FT}(x,x')$ are basically unknown, $A_{UT}^{\gamma {\rm SIDIS}}$ will be an important future measurement to learn about multi-parton correlations in the nucleon.

\end{abstract}

\maketitle

\renewcommand*{\thefootnote}{\arabic{footnote}}
\setcounter{footnote}{0}

\section{Introduction}
Multi-parton correlations in hadrons have been an area of intense interest for almost 50 years, starting with measurements of transverse single-spin asymmetries (TSSAs) in single-inclusive hadronic collisions back in the late 1970s~\cite{Bunce:1976yb,Klem:1976ui} and the theoretical work~\cite{Kane:1978nd,Efremov:1981sh,Efremov:1984ip,Qiu:1991pp,Qiu:1991wg,Qiu:1998ia} attempting to understand them.  Two matrix elements that show up frequently are the so-called dynamical twist-3 (quark-gluon-quark) functions that some denote as $F_{FT}(x,x')$ and $G_{FT}(x,x')$, where $x, x'$ are the momentum fractions carried by quarks in the nucleon that undergo a hard scattering, with the gluon in the interaction carrying momentum $(x-x')$.  Many processes have been calculated analytically that involve these functions, namely, TSSAs~\cite{Efremov:1981sh,Efremov:1984ip,Qiu:1991pp,Qiu:1991wg,Qiu:1998ia,Kanazawa:2000hz,Eguchi:2006qz,Kouvaris:2006zy,Eguchi:2006mc,Zhou:2008fb,Koike:2009ge,Vogelsang:2009pj,Metz:2012ui,Kang:2012ns,Schlegel:2012ve,Kanazawa:2013uia,Beppu:2013uda,Dai:2014ala,Gamberg:2014eia,Kanazawa:2015ajw,Yoshida:2016tfh,Chen:2016dnp,Chen:2017lvx,Rein:2025qhe,Rein:2025pwu} and longitudinal-transverse double-spin asymmetries (DSAs)~\cite{Kang:2011jw,Liang:2012rb,Metz:2012fq,Hatta:2013wsa,Kanazawa:2014tda,Kanazawa:2015ajw} in electron-nucleon and proton-proton collisions.  However, in all cases, $F_{FT}(x,x')$ and $G_{FT}(x,x')$ enter the cross sections under an integral over one of the momentum fractions or at a particular kinematic point (e.g., $x=x'$), which limits the information one can extract on these functions.   

In Ref.~\cite{Albaltan:2019cyc}, TSSAs for the semi-inclusive deep-inelastic production of isolated photons from lepton-nucleon scattering, denoted here as $A_{UT}^{\gamma {\rm SIDIS}}$, were analytically calculated.  The authors found this observable had a unique feature:~$F_{FT}(x,x')$ and $G_{FT}(x,x')$ enter point-by-point in their full support $x,x'$ at leading order in perturbative QCD~\cite{Albaltan:2019cyc}.  Therefore,  if $A_{UT}^{\gamma {\rm SIDIS}}$ is measurable at the Electron-Ion Collider (EIC), it could offer unprecedented insight into multi-parton correlations in the nucleon.\footnote{We also refer the reader to numerical studies in Refs.~\cite{Bauer:2022mvl,Fitzgibbons:2024zsh} on single-inclusive electron-nucleon TSSA and DSA observables at the EIC that are sensitive to quark-gluon-quark effects in hadrons.}  These functions are also needed for the evolution and operator product expansion of transverse momentum dependent (TMD) parton distribution functions (PDFs) like the Sivers and worm-gear functions~\cite{Kang:2008ey,Braun:2009mi,Zhou:2008mz,Ma:2012xn,Aybat:2011ge,Kanazawa:2015ajw,Scimemi:2018mmi,Scimemi:2019gge,Rein:2022odl} as well as in next-to-leading order results for TSSAs~\cite{Vogelsang:2009pj,Kang:2012ns,Dai:2014ala,Yoshida:2016tfh,Chen:2016dnp,Chen:2017lvx,Rein:2025qhe,Rein:2025pwu}.

To this end, we provide numerical estimates for $A_{UT}^{\gamma {\rm SIDIS}}$ for EIC kinematics, exploring a broad phase space in order to isolate the regions where the asymmetry could be sizable.  We utilize realistic models for $F_{FT}(x,x')$ and $G_{FT}(x,x')$ based on the first transverse momentum moment of the Sivers function and lattice QCD computations of the $d_2$ matrix element.  We also created a user-friendly google colab notebook~\cite{AUTgamSIDIS_lib} to allow the reader to explore other scenarios/assumptions for the quark-gluon-quark functions and experimental configurations.  The paper is organized as follows.  In Sec.~\ref{s:theory}, we review the theoretical formalism for $A_{UT}^{\gamma {\rm SIDIS}}$, including a re-writing of the final expression in Ref.~\cite{Albaltan:2019cyc} in a significantly more compact form.  In Sec.~\ref{s:model}, we put forth our models for $F_{FT}(x,x')$ and $G_{FT}(x,x')$, focusing on two different scenarios for the functions.  In Sec.~\ref{s:pred}, we show our results for $A_{UT}^{\gamma {\rm SIDIS}}$ in the form of multi-dimensional ``heat map'' plots to highlight where the asymmetry is largest.  We give concluding remarks in Sec.~\ref{s:concl}.

\vspace{-0.15cm}
\section{Theoretical Framework} \label{s:theory}
The process under consideration is the production of an isolated real photon from the hard scattering of a lepton on a nucleon, $\ell(l) + N(P) \to \ell(l') + \gamma(P_\gamma)+X$, where the momenta of the various particles are indicated in parentheses.\footnote{Even though the theoretical framework is generically for a nucleon, numerical results in this paper are implicitly for a proton.}  We also define $q = l-l'-P_\gamma$ and $\tilde{q} = l-l'$.  We work in the lepton-nucleon center-of-mass (CM) frame with the nucleon moving along $+z$ and the lepton moving along $-z$.  We will give numerical estimates for the asymmetry $A_{UT}^{\gamma {\rm SIDIS}}$ that occurs when the lepton is unpolarized and the nucleon is transversely polarized with spin $\vec{S}_T$ that we take to point along the $+y$ axis.  Some relevant kinematic variables are $s = (P+l)^2$, $Q^2=-q^2$, $\tilde{Q}^2=-\tilde{q}^2$, $x_B = Q^2/(2P\!\cdot\!q)$, and $\tilde{x}_B = \tilde{Q}^2/(2P\!\cdot\! \tilde{q})$.  The rapidity and transverse momentum of the scattered electron (produced photon), respectively, are  $\eta'$ ($\eta^\gamma$) and $\vec{p}^{\,\prime}_T$ ($\vec{p}_T^{\,\gamma}$), with $\phi'$ ($\phi^\gamma$) being the azimuthal angle of the latter.

Since there is no uniform notation in the literature for the quark-gluon-quark correlators under consideration in Eq.~\eqref{e:sigmaUT} below, to be definitive we provide our operator definitions of $F_{FT}(x,x')$ and $G_{FT}(x,x')$~\cite{Kanazawa:2015ajw}\footnote{Some other common definitions can be found in Refs.~\cite{Braun:2009mi,Kang:2008ey,Kanazawa:2010au}.  The reader can refer to Eqs.~(111), (112) of Ref.~\cite{Braun:2011aw} for the relation between the functions in Ref.~\cite{Braun:2009mi} and those in Refs.~\cite{Kang:2008ey,Kanazawa:2010au}.  For our functions in Eq.~\eqref{e:paramPhiF}, the relation to the functions in Ref.~\cite{Braun:2009mi} are $F_{FT}^q(x,x')=-\frac{1}{M}T_{\bar{q}F q}(-x',x'-x,x)$ and $G_{FT}^q(x,x')=-\frac{1}{M}\Delta T_{\bar{q}F q}(-x',x'-x,x)$.  (Note Ref.~\cite{Braun:2009mi} uses $\epsilon_{0123}=1$.)}:
\begin{align}
\Phi_{F}^{q,\,\rho}(x,x') &= \frac{M}{2}\Bigg(\epsilon^{Pn\rho S}\,\slash P\, iF_{FT}^{q}(x,x')-(S^{\rho}-P^{\rho}(n\cdot S))\,\slash P\gamma_{5}\, G_{FT}^{q}(x,x')+\dots\Bigg)\,,\label{e:paramPhiF}
\end{align}
where
\begin{align}
\Phi^{q,\,\rho}_{F,ij}(x,x') &= \int_{-\infty}^\infty\frac{d\lambda}{2\pi}\int_{-\infty}^\infty\frac{d\mu}{2\pi}\,\mathrm{e}^{ix' \lambda+i(x-x')\mu}\langle P,S|\,\bar{\psi}^q_j(0)\,\mathcal{W}[0\,;\mu n]\,ign_{\eta}F^{\eta \rho}(\mu n)\,\mathcal{W}[\mu n\,;\lambda n]\,\psi^q_i(\lambda n)\,|P,S\rangle\,. \label{e:PhiFxxp}
\end{align}
The ellipsis in Eq.~\eqref{e:paramPhiF} indicates additional dynamical twist-3 PDFs that do not enter the $A_{UT}^{\gamma {\rm SIDIS}}$ observable.  The vector $n$ is a lightcone vector conjugate to the nucleon momentum $P$ (i.e., $P\cdot n =1$ with $n^2=0$), and $S$ is the spin vector of the nucleon ($S^2=-1$).  In the matrix element \eqref{e:PhiFxxp}, $\psi^q$ is a quark field of flavor $q$, $F^{\mu\nu}$ is the field strength tensor, and $\mathcal{W}$ is a Wilson line connecting the respective parton fields.  Our convention for the Levi-Civita tensor is $\epsilon^{0123} = 1$.

The unpolarized cross section $d\sigma_{UU}$ in the denominator of $A_{UT}^{\gamma {\rm SIDIS}}$ was first calculated in the parton model in Ref.~\cite{Brodsky:1972} and then reproduced in Ref.~\cite{Albaltan:2019cyc}:
\begin{equation}
\frac{d\sigma_{UU}}{dp'_Td\eta'd\phi' dp_T^\gamma d\eta^\gamma d\phi^\gamma} = \frac{\alpha_{em}^3}{4\pi^2 s Q^4}p'_Tp_T^\gamma \!\sum_{n={\rm BH,C,I}}\hat{\sigma}_{UU}^n\,f_1^n(x_B,\mu_n)\,, \label{e:unpolcs}
\end{equation}
where $\alpha_{em}=e^2/4\pi$ is the electromagnetic coupling, $f_1(x,\mu)$ is the unpolarized PDF evaluated at momentum fraction $x$ and scale $\mu$, and $\hat{\sigma}^n_{UU}$ are hard factors that can be found in Appendix~A.1 of Ref.~\cite{Albaltan:2019cyc}. For our numerical computations, we use the central curve of $f_1$ from CT18NLO~\cite{Hou:2019qau}.  We note that there are three types of terms in Eq.~\eqref{e:unpolcs} called Bethe-Heitler (BH), Compton (C), and Interference (I) that are distinguished by if the cross section diagram has the final-state photon radiated from the lepton, from the quark, or an interference between those two~\cite{Albaltan:2019cyc}.  For a generic function $f$, we have
\begin{equation}
    f^{\rm BH} \equiv \sum_{q=u,d,s} e_q^2\left(f^q+f^{\bar{q}}\right)\,,\quad
f^{\rm C} \equiv \sum_{q=u,d,s} e_q^4\left(f^q+f^{\bar{q}}\right)\,,\quad
f^{\rm I} \equiv \sum_{q=u,d,s} e_q^3\left(f^q-f^{\bar{q}}\right), \label{e:BHCI}
\end{equation}
where $e_q$ is the fractional quark charge for flavor $q$, and the sum is only over light quarks $u,d,s$.  The scale at which $f$ is evolved to is chosen differently for each term:~$\mu_{\rm BH} = Q, \mu_{\rm C} = \tilde{Q}, \mu_{\rm I} = \sqrt{Q\tilde{Q}}$~\cite{Albaltan:2019cyc}.  A partonic description of this process is valid if $Q^2 \gg M^2, \tilde{Q}^2 \gg M^2$, and $Q^2-\tilde{Q}^2 \gg M^2$, where $M$ is the nucleon mass~\cite{Brodsky:1972}.

The transversely polarized cross section $d\sigma_{UT}$ in the numerator of $A_{UT}^{\gamma {\rm SIDIS}}$ was first computed in Ref.~\cite{Albaltan:2019cyc} to leading order in perturbative QCD.  The form of the cross section we report here, though, utilizes two observations that make the expression significantly more compact.  First, rather than the four azimuthal spin correlations $\left(\epsilon^{Pll'S},\epsilon^{PlP_\gamma S},l^\prime_T\cdot S_T\,\epsilon^{Pll'P_\gamma},P_{\gamma T}\cdot S_T\,\epsilon^{Pll'P_\gamma}\right)$ that show up in Eq.~(21) of Ref.~\cite{Albaltan:2019cyc}, the identity $g^{\alpha\beta}\epsilon^{\mu\nu\rho\sigma}=g^{\mu\beta}\epsilon^{\alpha\nu\rho\sigma}+g^{\nu\beta}\epsilon^{\mu\alpha\rho\sigma}+g^{\rho\beta}\epsilon^{\mu\nu\alpha\sigma}+g^{\sigma\beta}\epsilon^{\mu\nu\rho\alpha}$ allows us to reduce them down to two structures:~$\epsilon^{Pll'S}\propto \sin(\phi_S -\phi')$ and $\epsilon^{PlP_\gamma S}\propto \sin(\phi_S-\phi^\gamma)$.  Second, the hard factors simplify considerably if we introduce the following linear combinations of $F_{FT}(x,x')$ and $G_{FT}(x,x')$:~$F_\pm^n(x,x') \equiv F_{FT}^n(x,x') \pm G_{FT}^n(x,x')$.  The final result reads, 
\begin{equation}
    \frac{d\sigma_{UT}}{dp'_Td\eta'd\phi' dp_T^\gamma d\eta^\gamma d\phi^\gamma} = \frac{\alpha_{em}^3M}{4\pi s Q^8}p'_Tp_T^\gamma \left[\epsilon^{Pll'S}\,\sigma_{UT}^{\phi'} + \epsilon^{PlP_\gamma S}\,\sigma_{UT}^{\phi^\gamma}\right]\,,\label{e:polcs}
\end{equation}
where
\begin{equation}
    \sigma_{UT}^{\phi'(\phi^\gamma)} = \sum_{n={\rm C,I}}\left[\hat{\sigma}^{n,\phi'(\phi^\gamma)}_{{\rm HP},+}\,F_{+}^n(x_B,\tilde{x}_B,\mu_n) + \hat{\sigma}^{n,\phi'(\phi^\gamma)}_{{\rm HP},-}\,F_{-}^n(x_B,\tilde{x}_B,\mu_n) + \hat{\sigma}^{n,\phi'(\phi^\gamma)}_{{\rm SFP},+}\,F_{+}^n(x_B,0,\mu_n) + \hat{\sigma}^{n,\phi'(\phi^\gamma)}_{{\rm SFP},-}\,F_{-}^n(x_B,0,\mu_n)\right].\label{e:sigmaUT}
\end{equation}
The partonic cross sections $\hat{\sigma}^{n,\phi'(\phi^\gamma)}_{{\rm HP},+}$, $\hat{\sigma}^{n,\phi'(\phi^\gamma)}_{{\rm HP},-}$, $\hat{\sigma}^{n,\phi'(\phi^\gamma)}_{{\rm SFP},+}$, $\hat{\sigma}^{n,\phi'(\phi^\gamma)}_{{\rm SFP},-}$ (for $n={\rm C, I}$) associated with hard pole (HP) and soft-fermion pole (SFP) matrix elements are given here in~\ref{s:partCS}.  The exact formulas for $\epsilon^{Pll'S}$ and $\epsilon^{PlP_\gamma S}$ are also provided in Eq.~\eqref{e:azimuthal}. Note, as was already discussed in Ref.~\cite{Albaltan:2019cyc}, that in the transversely polarized cross section, the BH term as well as the soft-gluon pole (SGP) correlator pieces cancel out between contributions from different diagrams.

We define the asymmetry $A_{UT}^{\gamma {\rm SIDIS}}$ as
\begin{equation}
    A_{UT}^{\gamma {\rm SIDIS}}(p'_T,\eta',\phi' ,p_T^\gamma ,\eta^\gamma ,\phi^\gamma)\equiv \frac{\frac{1}{2}[d\sigma_{UT}(\uparrow)-d\sigma_{UT}(\downarrow)]}{d\sigma_{UU}}=\frac{d\sigma_{UT}(\phi_s\!=\!\frac{\pi}{2})}{d\sigma_{UU}}\,, \label{e:AUT}
\end{equation}
where the numerator and denominator are given in Eqs.~\eqref{e:unpolcs}, \eqref{e:polcs}, respectively, and $\uparrow (\downarrow)$ denotes the nucleon  spin along the $\pm y$ axis.  We make explicit the dependence of $A_{UT}^{\gamma {\rm SIDIS}}$ on the six variables $(p'_T,\eta',\phi' ,p_T^\gamma ,\eta^\gamma ,\phi^\gamma)$.  In Sec.~\ref{s:pred} we explore the asymmetry in this phase space in an attempt to isolate in which regions it could be measurable at the EIC.

\vspace{-0.15cm}
\section{Model Input for Quark-Gluon-Quark Correlators} \label{s:model}

The quark-gluon-quark functions $F_{FT}(x,x')$ and $G_{FT}(x,x')$ are basically unknown.  The advantage of $A_{UT}^{\gamma {\rm SIDIS}}$ is the ability to probe these correlators point-by-point in their full support $x,x'$, which is an unprecedented feature of this observable.  If the asymmetry is sizable enough to be measured at the EIC, then first-of-its-kind information could be obtained on multi-parton correlations in the nucleon.  In order to provide such estimates, a reasonable model must be developed for $F_{FT}(x,x')$ and $G_{FT}(x,x')$.  Such an undertaking was already disseminated in Appendix~E of Ref.~\cite{Rein:2025qhe}.  We recap the important features here.

First, let us recall some properties of $F_{FT}(x,x')$ and $G_{FT}(x,x')$.  The support for these functions lies in the region $|x|<1, |x'|<1, |x-x'|<1$, and they satisfy the following relations (see, e.g., Ref.~\cite{Kanazawa:2015ajw}):
\begin{align}
    F_{FT}(x,x') = F_{FT}(x',x)\,, &\quad G_{FT}(x,x') = -G_{FT}(x',x)\,, \label{e:sym}\\
    F_{FT}^{\bar{q}}(x,x') = F_{FT}^q(-x',-x)\,, &\quad G_{FT}^{\bar{q}}(x,x') = -G_{FT}^q(-x',-x)\,. \label{e:antiq}
\end{align}
The SGP matrix element $F_{FT}(x,x)$ is also connected to the first transverse momentum moment of the Sivers function~\cite{Boer:2003cm,Kanazawa:2015ajw}:
\begin{equation}
    \pi\,F_{FT}^q(x,x) = \int \!d^2\vec{k}_T\, \frac{\vec{k}_T^2}{2M^2}\,f_{1T}^{\perp,q}(x,\vec{k}_T)\equiv f_{1T}^{\perp(1),q}(x)\,. \label{e:sivers}
\end{equation}
Second, we make a change of coordinates from $(x,x')$ to $(r,\phi)$, where $x = r\cos(\phi+\frac{\pi}{4})$ and $x'=r\sin(\phi+\frac{\pi}{4})$, with 
\begin{equation}
    r=\sqrt{x^2+(x')^2}\,,\quad\quad
\phi = \begin{cases}
-\frac{\pi}{4}+\arctan(x'/x)\;{\rm if}\; x\geq 0, x'\geq x\\
\;\frac{3\pi}{4}+\arctan(x'/x)\;{\rm if}\; x< 0\\
\;\frac{7\pi}{4}+\arctan(x'/x)\;{\rm if}\; x\geq 0, x' < x\,.
\end{cases}
\end{equation}
The polar coordinates have been chosen so that we count $\phi$ starting from the ``diagonal'' axis of support ($x'=x$) instead of from the $x$-axis.  This is convenient since at $\phi=0$, one has $F_{FT}(x,x)=f_{1T}^{\perp(1)}(x)/\pi$ and $G_{FT}(x,x)=0$ (see Eqs.~\eqref{e:sym}, \eqref{e:sivers}).  Lastly, we expand the $\phi$ dependence of $F_{FT}(x,x')$ and $G_{FT}(x,x')$ in a Fourier series of the following form (see Appendix~E of Ref.~\cite{Rein:2025qhe} for more details):
\begin{align}
F_{FT}^q(x,x')\big|_{\rm model} &= \left\{\frac{1}{2\pi}f_{1T}^{\perp(1),q+\bar{q}}(\tfrac{r}{\sqrt{2}})\left[1+\sum_{n=1}^3[a_{2n}^q (\cos(2n\phi)-1)]\right]\right.\nonumber \\
&\hspace{0.5cm}\left.+\frac{1}{2\pi}f_{1T}^{\perp(1),q-\bar{q}}(\tfrac{r}{\sqrt{2}})\left[\cos(\phi)+\sum_{n=1}^3[a_{2n+1}^q(\cos((2n+1)\phi)-\cos(\phi))]\right]\right\}e(x,x')\,, \label{e:FFTmodel} \\[0.3cm]
G_{FT}^q(x,x')\big|_{\rm model} &= \left\{-\frac{1}{\pi}f_{1T}^{\perp(1),q+\bar{q}}(\tfrac{r}{\sqrt{2}})\sum_{n=1}^6[b_n^q\sin(n\phi)]\right\}e(x,x')\,,\label{e:GFTmodel}
\end{align}
where $f_{1T}^{\perp q\pm \bar{q}}\equiv f_{1T}^{\perp,q}\pm f_{1T}^{\perp,\bar{q}}$, and $e(x,x')$ is an ``enveloping'' function that allows $F_{FT}^q(x,x')$ and $G_{FT}^q(x,x')$ to remain smooth at the boundaries $|x|, |x'|,\,{\rm or}\, |x-x'| \to 1$:  
\begin{align}
    e(x,x')&=\left(\frac{2}{1+e^{-50(1-x^2)^3}}-1\right)\left(\frac{2}{1+e^{-50(1-(x')^2)^3}}-1\right)\left(\frac{2}{1+e^{-50(1-(x-x')^2)^3}}-1\right)\theta(1-|x|)\,\theta(1-|x'|)\,\theta(1-|x-x'|)\,.
\end{align}
We note the relations~\eqref{e:sym}, \eqref{e:sivers} are respected by \eqref{e:FFTmodel}, \eqref{e:GFTmodel}.
In writing down Eqs.~\eqref{e:FFTmodel}, \eqref{e:GFTmodel}, we also made the following assumptions:~(i)~all Fourier coefficients $a_n$ and $b_n$ are constants (independent of $r$); (ii)~all Fourier coefficients $a_n$ for $n \ge 8$ and $b_n$ for $n \ge 7$ vanish; (iii)~$f_{1T}^{\perp(1)}(x)$ sets the size for $F_{FT}(x,x')$ and $G_{FT}(x,x')$, i.e., it serves as the overall normalization to the quark-gluon-quark functions. We therefore have the following unknown coefficients that need to be determined in order to make numerical estimates for $A_{UT}^{\gamma\,{\rm SIDIS}}$:
\begin{equation}
    \boldsymbol{a^q} \equiv (a_2^q,a_4^q,a_6^q;a_3^q,a_5^q,a_7^q)\,,\quad\quad
\boldsymbol{b^q} \equiv (b_1^q,b_2^q,b_3^q,b_4^q,b_5^q,b_6^q)\,.
\end{equation}
(Note the ordering of the $a$ coefficients, where the even ones are listed first and then the odd.)  

\begin{figure}[t!]
\begin{center}
\includegraphics[width=0.85\textwidth]{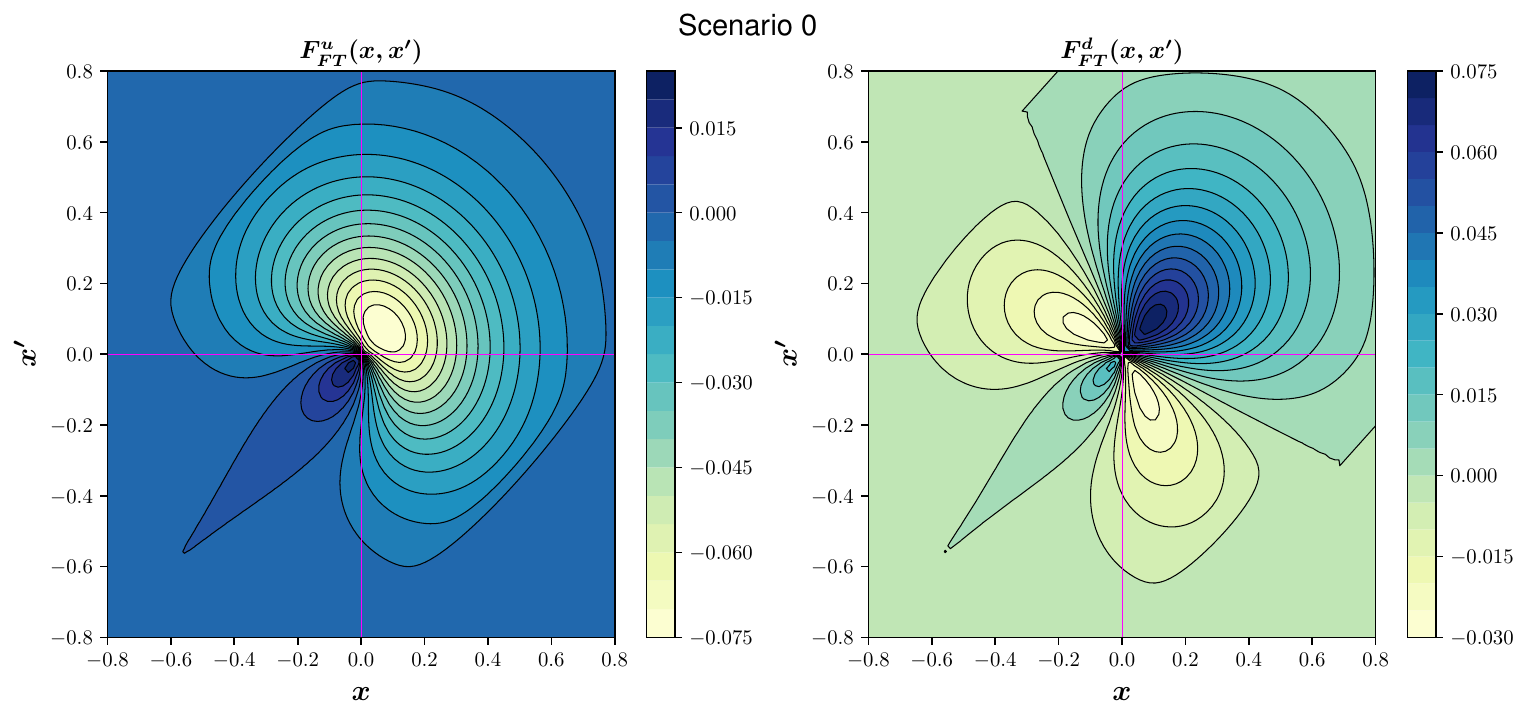}\vspace{-0.5cm}
\end{center}
\caption{$F_{FT}(x,x')$ vs.~$(x,x')$ at a scale $\mu^2=4\,{\rm GeV^2}$ for Scenario~0 for the up quark (left) and down quark (right) in a proton.  Recall that $G_{FT}(x,x')=0$ for Scenario~0.\vspace{-0.3cm}} 
\label{f:FFT0}
\end{figure}
\begin{figure}[t!]
\begin{center}
\includegraphics[width=0.85\textwidth]{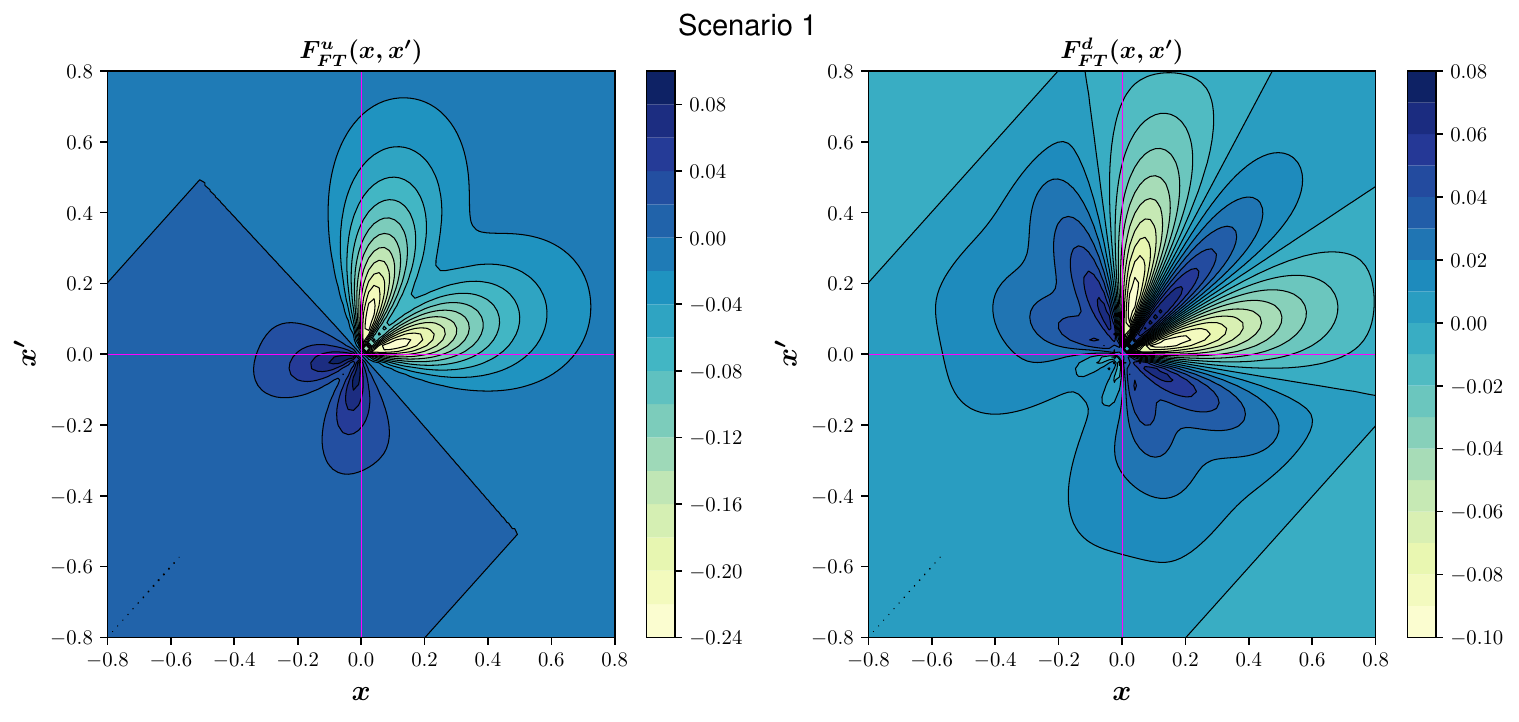}
\includegraphics[width=0.85\textwidth]{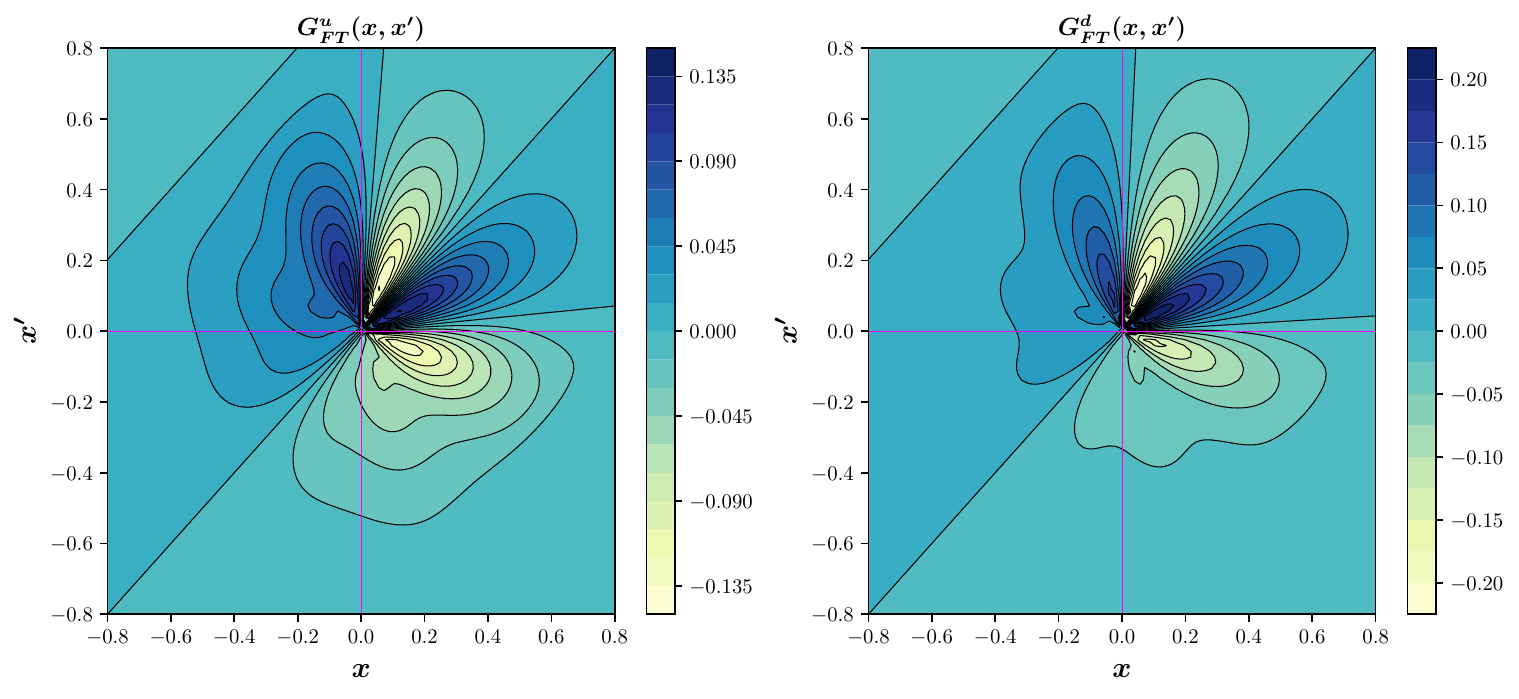}\vspace{-0.5cm}
\end{center}
\caption{$F_{FT}(x,x')$ vs.~$(x,x')$ (top row) and $G_{FT}(x,x')$ vs.~$(x,x')$ (bottom row) at a scale $\mu^2=4\,{\rm GeV^2}$ for Scenario~1 for the up quark (left) and down quark (right) in a proton.\vspace{-0.3cm}} 
\label{f:FFT_GFT1}
\end{figure}

Another piece of information we can utilize to help create a reasonable model is lattice QCD calculations of the $d_2$ matrix element~\cite{Gockeler:2005vw,Bickerton:2020hjo,Burger:2021knd,Crawford:2024wzx}, which is connected to the color Lorentz force mediated by the strong force in the nucleon~\cite{Burkardt:2008ps,Aslan:2019jis} and can be expressed as~\cite{Burkardt:2008ps,Aslan:2019jis} (see also Refs.~\cite{Shuryak:1981pi,Jaffe:1989xx}) 
\begin{equation}
    d_2^q = -\int_{-1}^1 dx\int_{-1}^1 dx'\,F_{FT}^q(x,x')\,.
\end{equation}
This allows us to relate one of the unknown coefficients in $F_{FT}(x,x')$, say $a_2$, to $a_{4,6,3,5,7}$ once they are chosen:
\begin{equation}
    a_2^q = (d_2^q-(A_0^q + A_4^q a_4^q + A_6^q a_6^q + A_3^q a_3^q + A_5^q a_5^q + A_7^q a_7^q))/A_2^q\,, \label{e:a2}
\end{equation}
where
\begin{equation}
   A_0^q \equiv -\int_{-1}^1 dx \int_{-1}^1 dx'\,F_{FT}^q(x,x')\bigg |_{\boldsymbol{a^q}=0}\,,\quad\quad
A_i^q \equiv -A_0^q -\int_{-1}^1 dx \int_{-1}^1 dx'\,F_{FT}^q(x,x')\bigg |_{a_i^q=1,a_{j\neq i}^q=0}\,. \label{e:Ai}
\end{equation}  

Now that we have set up our model for the quark-gluon-quark correlators, we list two scenarios below for $\boldsymbol{a^q}, \boldsymbol{b^q}$ that will be implemented for our $A_{UT}^{\gamma {\rm SIDIS}}$ numerical estimates.  Figures~\ref{f:FFT0} and \ref{f:FFT_GFT1} show the resulting $F^{u,d}_{FT}(x,x')$ and $G^{u,d}_{FT}(x,x')$ from these two scenarios.  For both of these, we use the $d_2$ lattice (central) values from Ref.~\cite{Burger:2021knd}, which are given by $d_2^{u} = 0.026(4)(13), d_2^{d} = -0.0086(26)(146)$.  Even though there are large uncertainties on the lattice calculations of $d_2$, along with limitations related to renormalization and unphysical pion masses, we still find these are useful in helping us formulate reasonable output for $F_{FT}(x,x')$ and $G_{FT}(x,x')$.  The strange quark $a$'s and $b$'s are set to zero in both scenarios.  We use the central curves of the JAM3D-22 extraction of $f_{1T}^{\perp(1),q}(x)$~\cite{Gamberg:2022kdb}, specifically the one where (up and down) antiquarks are included (see Sec.~IV of Ref.~\cite{Gamberg:2022kdb}), i.e., $q=u,d,\bar{u},\bar{d}$.  The initial scale for our model is $\mu_0^2 = 2\,{\rm GeV^2}$, and the evolution of $F_{FT}(x,x')$ and $G_{FT}(x,x')$ is simply ``inherited'' from the evolution of $f_{1T}^{\perp(1)}(x)$ in Eqs.~\eqref{e:FFTmodel}, \eqref{e:GFTmodel}, which is sufficient for our purposes.\footnote{In JAM3D-22~\cite{Gamberg:2022kdb}, a DGLAP-type evolution for the Sivers function was used, analogous to Ref.~\cite{Duke:1983gd}, where a double-logarithmic $Q^2$-dependent term was explicitly added to the parameters in an attempt to mimic the full twist-3 evolution. Any mixing with tri-gluon correlators or off-diagonal quark-gluon-quark matrix elements in the evolution was ignored.}.  We mention that numerical code for the full twist-3 evolution of quark-gluon-quark correlators can be found in Ref.~\cite{Rodini:2024usc}.  While we do not use that code here, we remark that in asymmetries, DGLAP-type evolution effects typically cancel out between the numerator and denominator.  For example, fits using DGLAP-type evolution~\cite{Anselmino:2016uie,Gamberg:2022kdb}  are able to equally well describe Sivers asymmetry data as those with full TMD evolution~\cite{Echevarria:2020hpy,Bacchetta:2020gko,Bury:2021sue} across a wide $Q^2$ range.   
\\

\noindent \underline{Scenario 0}: 
\begin{align}
    \boldsymbol{a^u} = (-0.2691,0,0;0,0,0)\,,&\quad \boldsymbol{a^d} = (0.7822,0,0;0,0,0)\,,\quad \boldsymbol{b^u} = (0,0,0,0,0,0)\,,\quad \boldsymbol{b^d} = (0,0,0,0,0,0)\nonumber
\end{align}
In this case all parameters are set to zero except for $a_2^{u(d)}$, which is then entirely fixed by the lattice value for $d_2^{u(d)}$.  Note that $G_{FT}(x,x')$ is identically zero in this scenario since all the $b$ coefficients have been set to zero.  We consider Scenario 0 as a ``minimalistic'' situation for $A_{UT}^{\gamma {\rm SIDIS}}$. 
\\

\noindent \underline{Scenario 1}: 
\begin{align}
   &\boldsymbol{a^u} = (1.1585,-\tfrac{2}{3},-\tfrac{2}{3};-\tfrac{1}{3},-1,-\tfrac{1}{3})\,,\quad
\boldsymbol{a^d} = (-0.6658,\tfrac{2}{3},\tfrac{2}{3};\tfrac{1}{3},1,\tfrac{1}{3})\,,\nonumber\\
&\boldsymbol{b^u} = (-1.1585,\tfrac{1}{3},\tfrac{2}{3},1,\tfrac{2}{3},\tfrac{1}{3})\,,\quad
\hspace{0.9cm} \boldsymbol{b^d} = (0.6658,-\tfrac{1}{3},-\tfrac{2}{3},-1,-\tfrac{2}{3},-\tfrac{1}{3})\nonumber 
\end{align}
In this case we have chosen $a^{u}_{4,6,3,5,7}$ and $b^{u}_{2,3,4,5,6}$ to be arbitrary values between $-1$ and $1$ and picked $a^{d}_{4,6,3,5,7}=-a^{u}_{4,6,3,5,7}$ and $b^{d}_{2,3,4,5,6}=-b^{u}_{2,3,4,5,6}$, so that none of the terms of the Fourier series \eqref{e:FFTmodel}, \eqref{e:GFTmodel} have a magnitude larger than $f_{1T}^{\perp(1)}(x)/\pi$.  We again emphasize that once $a^{u,d}_{4,6,3,5,7}$ are chosen, $a^{u,d}_2$ is determined by Eqs.~\eqref{e:a2}, \eqref{e:Ai}.  We also have fixed $b_1^{u,d}=-a_1^{u,d}$ in our model.

\vspace{-0.15cm}
\section{Numerical Estimates for the Electron-Ion Collider} \label{s:pred}
In this section we present selected results for $A_{UT}^{\gamma {\rm SIDIS}}$, defined in Eq.~\eqref{e:AUT}, at EIC kinematics. 
\begin{figure}[t!]
\begin{center}
\includegraphics[width=0.85\textwidth]{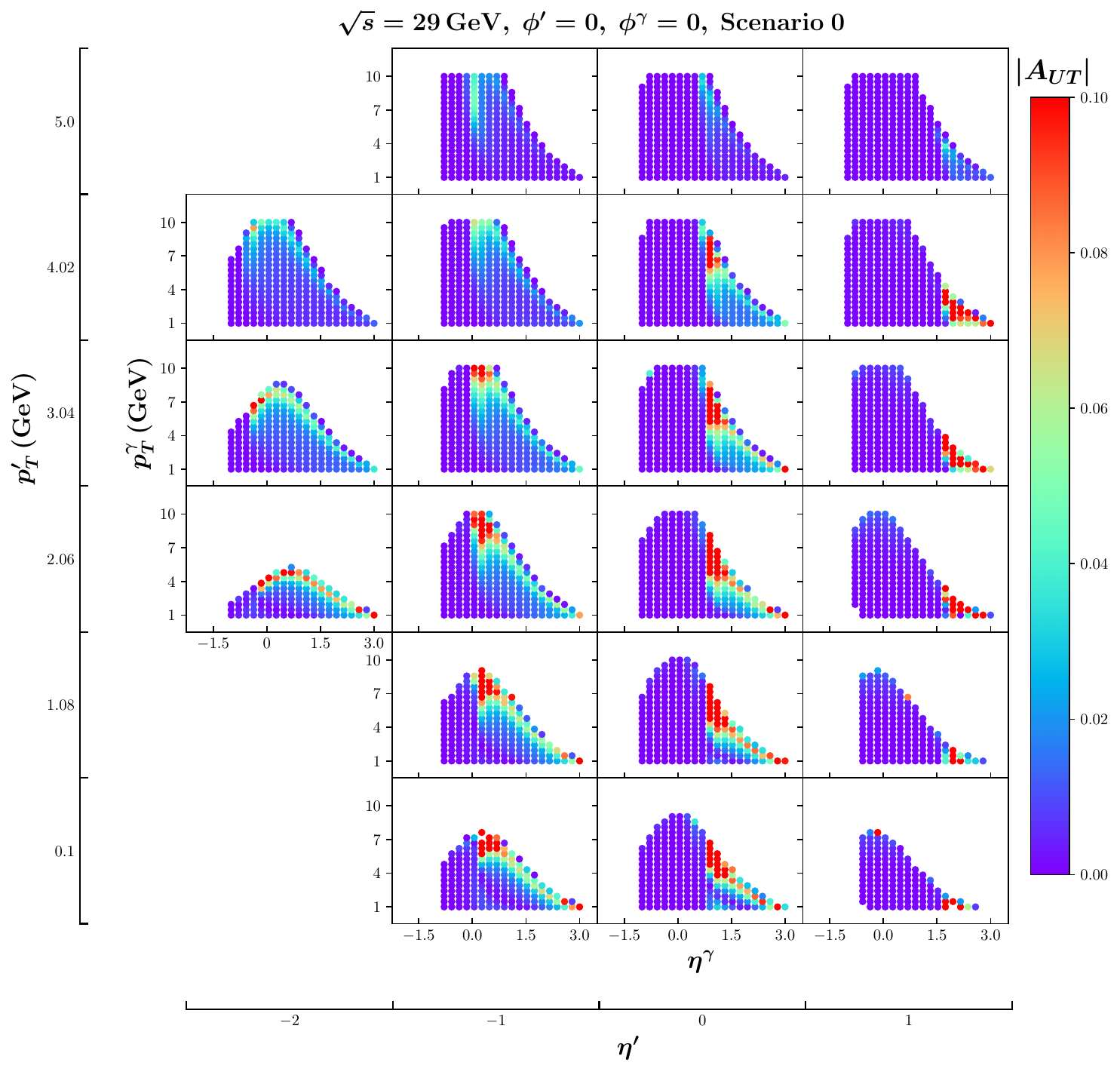}
\end{center}
\vspace{-0.5cm}
\caption{$\big|A_{UT}^{\gamma {\rm SIDIS}}\big|$ vs.~$(\eta',p_T^\prime,\eta^\gamma,p_T^\gamma)$ for $\sqrt{s}=29\,{\rm GeV}$ with $\phi'=\phi^\gamma=0$ for Scenario~0.  The outer axes are bins of $(\eta',p_T^\prime)$ and the inner axes are bins of $(\eta^\gamma,p_T^\gamma)$. The colors indicate the size of the asymmetry from close to zero (purple) to several percent (light blue/green) to at least 10\%~(red), with the exact mapping shown by the bar on the right side of the plot (note we have dropped the ``$\gamma {\rm SIDIS}$'' superscript on the $|A_{UT}|$ label for brevity).\vspace{-0.3cm}} 
\label{f:AUTscen0}
\end{figure}
\begin{figure}[t!]
\begin{center}
\includegraphics[width=0.85\textwidth]{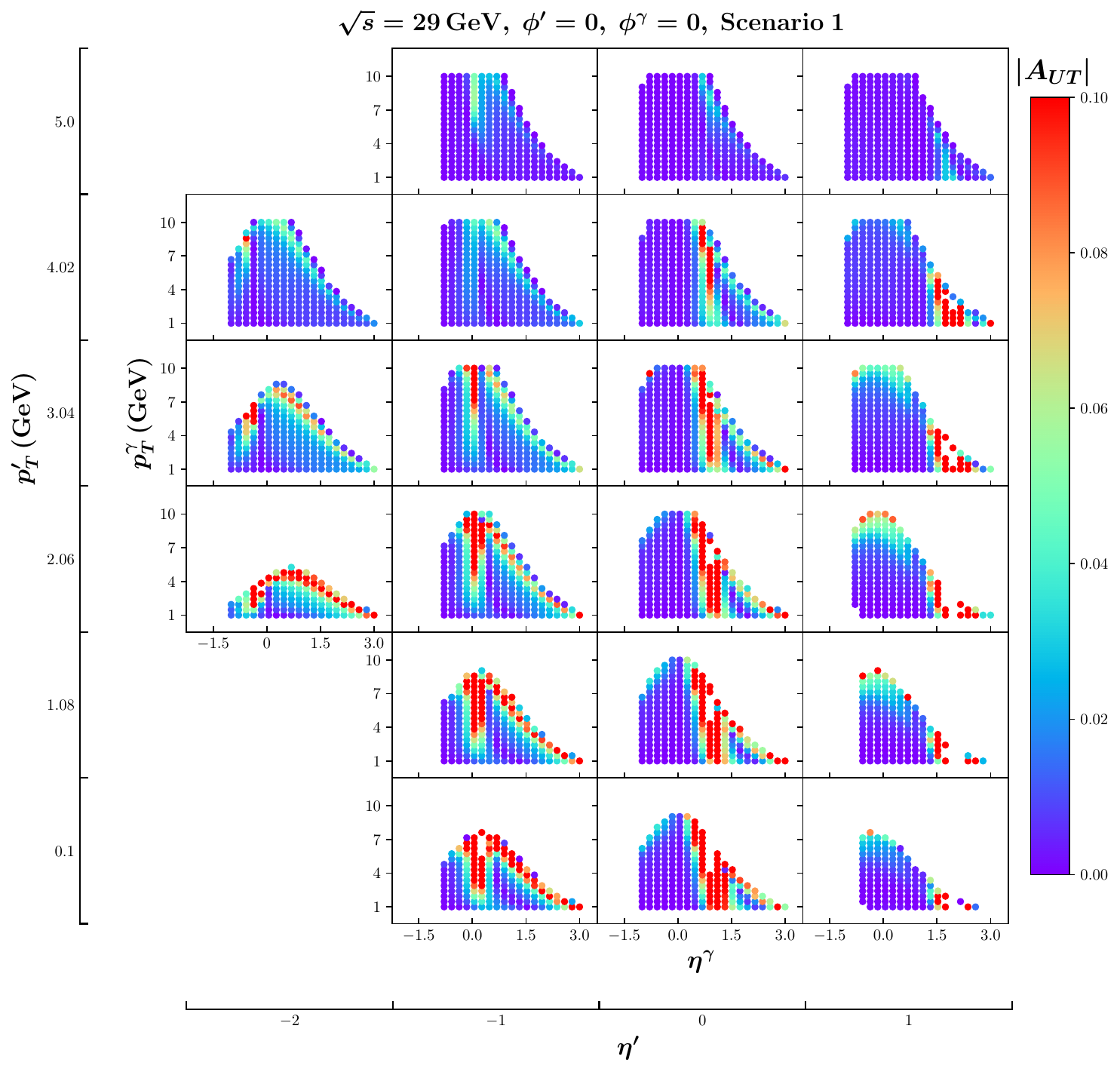}
\end{center}
\vspace{-0.5cm}
\caption{$\big|A_{UT}^{\gamma {\rm SIDIS}}\big|$ vs.~$(\eta',p_T^\prime,\eta^\gamma,p_T^\gamma)$ for $\sqrt{s}=29\,{\rm GeV}$ with $\phi'=\phi^\gamma=0$ for Scenario~1.  The outer axes are bins of $(\eta',p_T^\prime)$ and the inner axes are bins of $(\eta^\gamma,p_T^\gamma)$. The colors indicate the size of the asymmetry from close to zero (purple) to several percent (light blue/green) to at least 10\%~(red), with the exact mapping shown by the bar on the right side of the plot (note we have dropped the ``$\gamma {\rm SIDIS}$'' superscript on the $|A_{UT}|$ label for brevity).\vspace{-0.3cm}} 
\label{f:AUTscen1}
\end{figure}
\begin{figure}[h!]
\begin{center}
\includegraphics[width=0.48\textwidth]{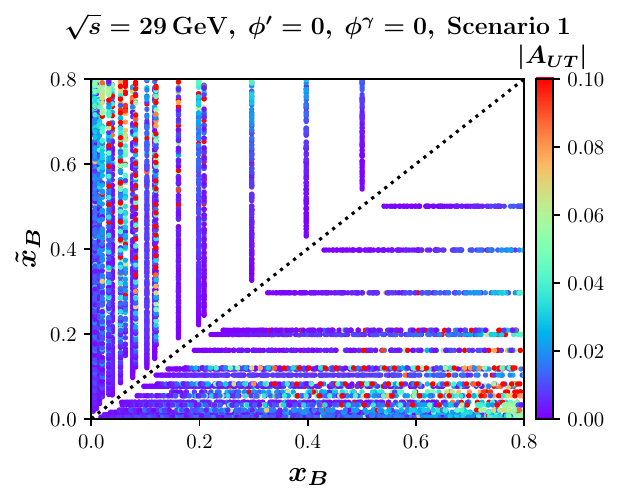}
\hspace{0.5cm}\includegraphics[width=0.48\textwidth]{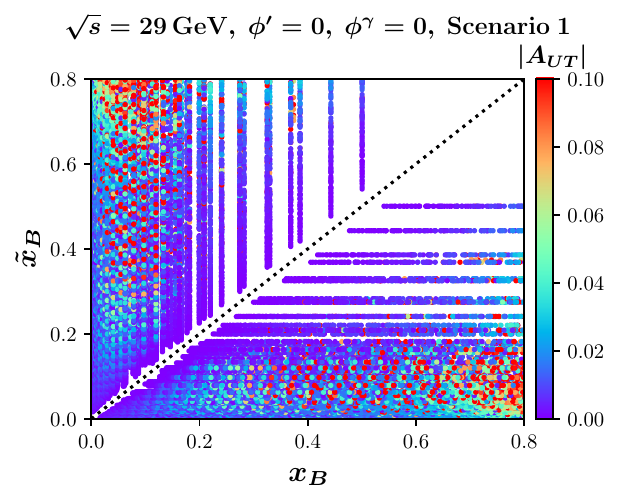}
\vspace{-1cm}
\end{center}
\caption{Plot of the kinematic points $\tilde{x}_B$ vs.~$x_B$ that enter the arguments of $F_{FT}$ and $G_{FT}$ corresponding to the setup in Fig.~\ref{f:AUTscen1} ($\sqrt{s}=29\,{\rm GeV}$ with $\phi'=\phi^\gamma=0$ for Scenario~1).  The colors indicate the magnitude of $A_{UT}^{\gamma {\rm SIDIS}}$ using the same scale as Fig.~\ref{f:AUTscen1}. In the left panel we show the binning from Fig.~\ref{f:AUTscen1}, where there are 4 bins of $\eta'$, 6 bins of $p_T^\prime$, and 20 bins each of $\eta^\gamma$ and $p_T^\gamma$.  In the right panel we show a finer binning in $\eta'$ and $p_T^\prime$, where each now has 10 bins.  The dashed black line indicates $x_B=\tilde{x}_B$.  Note that the experimental coverage only explicitly gives points below the $x_B=\tilde{x}_B$ line, but we have reflected the points across that line due to the symmetry properties~\eqref{e:sym}.  That is, information on $F_{FT}$ and $G_{FT}$ at $(x_B,\tilde{x}_B)$ also tells us about the point $(\tilde{x}_B,x_B)$.
\vspace{-0.3cm}} 
\label{f:xBxBtilde}
\end{figure}
\begin{figure}[h!]
\begin{center}
\includegraphics[width=0.85\textwidth]{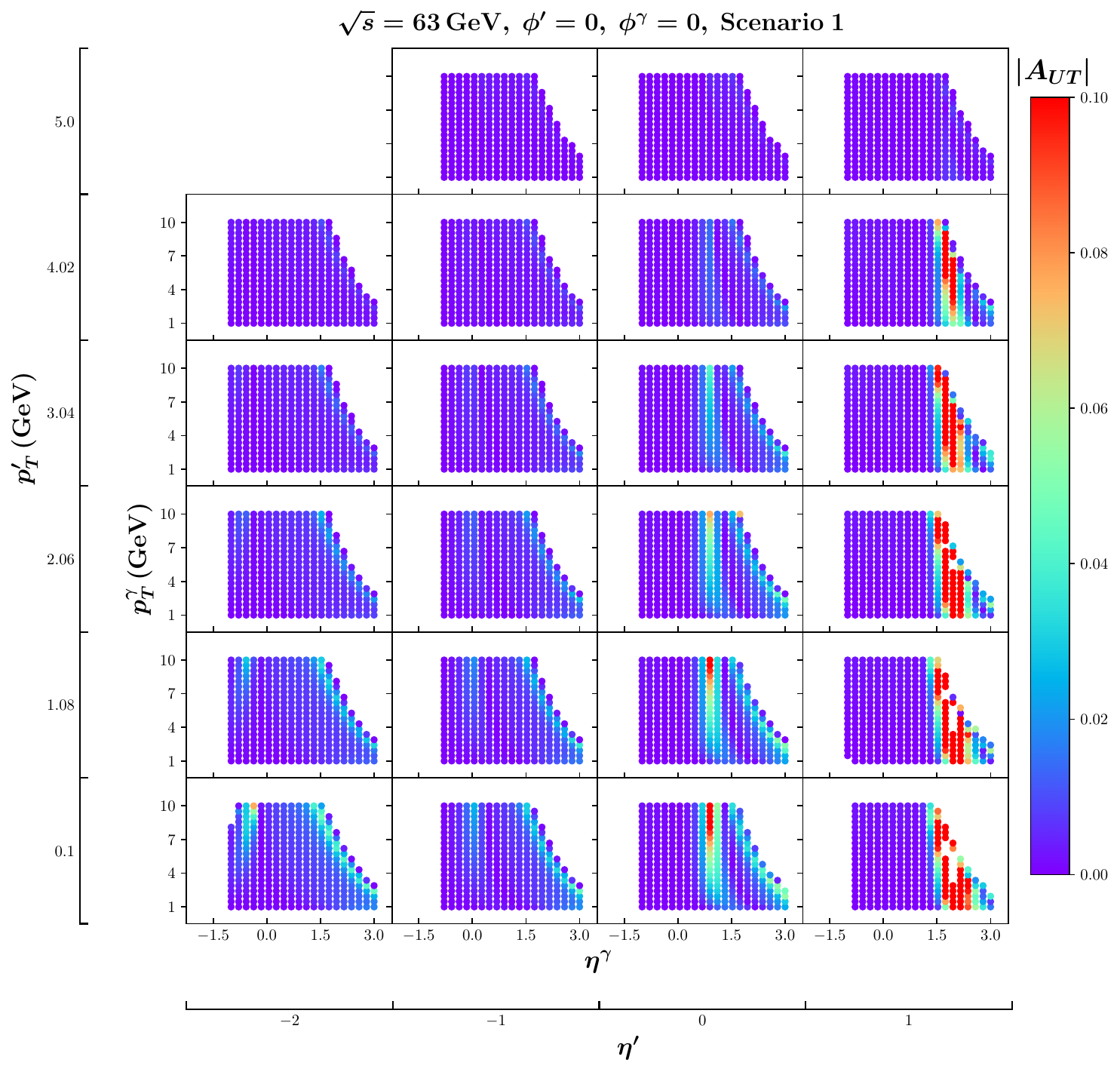}
\end{center}
\vspace{-0.5cm}
\caption{$\big|A_{UT}^{\gamma {\rm SIDIS}}\big|$ vs.~$(\eta',p_T^\prime,\eta^\gamma,p_T^\gamma)$ for $\sqrt{s}=63\,{\rm GeV}$ with $\phi'=\phi^\gamma=0$ for Scenario~1.  The outer axes are bins of $(\eta',p_T^\prime)$ and the inner axes are bins of $(\eta^\gamma,p_T^\gamma)$. The colors indicate the size of the asymmetry from close to zero (purple) to several percent (light blue/green) to at least 10\%~(red), with the exact mapping shown by the bar on the right side of the plot (note we have dropped the ``$\gamma {\rm SIDIS}$'' superscript on the $|A_{UT}|$ label for brevity).} 
\label{f:AUTscen1_63}
\end{figure}
Our goal is to sample a wide kinematic region and find generally where one can expect the asymmetry to have the largest magnitude, in order to guide future experiments.  For this reason, we will present multi-dimensional ``heat map'' plots showing $\big|A_{UT}^{\gamma {\rm SIDIS}}\big|$ in coarser bins of $(\eta',p_T^\prime)$ and finer bins of $(\eta^\gamma,p_T^\gamma)$ with $\phi'$ and $\phi^\gamma$ fixed.  A user-friendly google colab notebook is available~\cite{AUTgamSIDIS_lib} for the reader to explore other choices for model parameters for $F_{FT}(x,x')$ and $G_{FT}(x,x')$ as well as alternative experimental configurations.

We note that only kinematic points are kept that satisfy $Q^2>1\,{\rm GeV^2}, \tilde{Q}^2>1\,{\rm GeV^2}$, and $Q^2-\tilde{Q}^2>1\,{\rm GeV^2}$ and also do not cause $\big|A_{UT}^{\gamma {\rm SIDIS}}\big|>1$, as can sometimes happen at the edges of phase space where $x$ or $x'$ goes to 1. In the plots, a color mapping has been chosen so that purple regions indicate asymmetries close to zero, light bluish/greenish regions indicate asymmetries of several percent, and red regions indicate asymmetries of {\it at least} 10\%. As alluded to above, we must use caution when large asymmetries arise at the periphery of the subgraphs.  Instead, for each subgraph, we direct the reader's attention to the $\big|A_{UT}^{\gamma {\rm SIDIS}}\big|$ values given by the portion of points that lie away from the edges, which should provide the most reliable estimates.  

In Fig.~\ref{f:AUTscen0}, we show the results for CM energy $\sqrt{s}=29\,{\rm GeV}$ with $\phi'=\phi^\gamma=0$ for Scenario~0.  Recall that we consider this case to be a ``minimalistic'' scenario for $\big|A_{UT}^{\gamma {\rm SIDIS}}\big|$ since many of the quark-gluon-quark parameters (including all of the ones for $G_{FT}(x,x')$) are set to zero.  We see from the subgraphs in the plot that the asymmetry (ignoring the very edges of the kinematics) can be around $3$--$5\%$ at mid to backward rapidity of the outgoing electron ($\eta' = 0$ or $-1$), mid to forward rapidity of the photon ($\eta^\gamma \gtrsim 0$), smaller outgoing electron transverse momenta ($p_T^\prime \lesssim 3\,{\rm GeV}$), and larger photon transverse momenta ($p_T^\gamma \gtrsim 3\,{\rm GeV}$).

The same experimental configuration but now with Scenario~1, where all parameters for $F_{FT}(x,x')$ and $G_{FT}(x,x')$ are nonzero, is displayed in Fig.~\ref{f:AUTscen1}.  We see similar kinematic regions as were described previously give the largest $\big|A_{UT}^{\gamma {\rm SIDIS}}\big|$, but it is now $10\%$ or more.  There are also measurable asymmetries at smaller photon momentum ($p_T^\gamma \sim 1\,{\rm GeV}$), especially at mid-rapidity of the outgoing electron. In addition, we remark that it is preferable for both the outgoing electron and photon to be produced in the same direction (i.e.,  $\phi'=\phi^\gamma=0$ or $\phi'=\phi^\gamma=\pi$) in order to obtain the largest asymmetries.  We checked that if $\phi'=0$ and $\phi^\gamma=\pi$ (or vice versa), then the asymmetry drops down to $\sim \!\! 1$--$3\%$.  In addition, we confirmed that our statements about the phase space where $\big|A_{UT}^{\gamma {\rm SIDIS}}\big|$ is the largest do not qualitatively change if we use different values for $d_2^u$ and $d_2^d$ (namely, we picked $d_2^u = -0.00365,\, d_2^d = 0$ based on Ref.~\cite{Bickerton:2020hjo}) or a different extraction of the Sivers function (namely, from Ref.~\cite{Anselmino:2008sga}, which has a larger magnitude and different $x$ behavior than JAM3D-22). 

In Fig.~\ref{f:xBxBtilde}, we display the values of $x_B$ and $\tilde{x}_B$ that will be covered by the experimental setup in Fig.~\ref{f:AUTscen1}.  We show results for both the binning in Fig.~\ref{f:AUTscen1} as well as a finer binning in $\eta'$ and $p_T^\prime$ (see the caption of Fig.~\ref{f:xBxBtilde} for details).  The colors indicate the magnitude of $A_{UT}^{\gamma {\rm SIDIS}}$ using the same scale as Fig.~\ref{f:AUTscen1}.  This allows us to identify the regions of $(x,x')$ for $F_{FT}$ and $G_{FT}$ to which the observable will be most sensitive.  We see, from where the asymmetry has the largest magnitude, that these kinematic points are $(x,x')\in ([0.2,0.8],[0,0.1])$ and $(x,x')\in ([0,0.1],[0.2,0.8])$.  A caveat to this direct probe of $F_{FT}$ and $G_{FT}$ at those off-diagonal momentum fractions is that the transversely polarized cross section~(Eqs.~\eqref{e:polcs}, \eqref{e:sigmaUT}) depends not only on the HP structure $F_{FT}(x_B,\tilde{x}_B)\pm G_{FT}(x_B,\tilde{x}_B)$ but also the SFP $F_{FT}(x_B,0)\pm G_{FT}(x_B,0)$. We found that in most of the phase space where $\big|A_{UT}^{\gamma {\rm SIDIS}}\big|$ is sizable, the SFP contribution is comparable or larger than the HP, which has the potential to ``dilute'' the signal from $F_{FT}(x_B,\tilde{x}_B)$ and $G_{FT}(x_B,\tilde{x}_B)$.  On the other hand, the SFP functions $F_{FT}(x,0)$ and $G_{FT}(x,0)$ have never been extracted from experimental data either, so sensitivity to them from $A_{UT}^{\gamma {\rm SIDIS}}$ is also valuable.

In Fig.~\ref{f:AUTscen1_63}, we demonstrate what happens to the asymmetry at larger EIC CM energies, in this case $\sqrt{s}=63\,{\rm GeV}$, using Scenario~1.  There is a dramatic decrease in $\big|A_{UT}^{\gamma {\rm SIDIS}}\big|$ to basically zero everywhere except at mid to forward rapidity of the outgoing electron ($\eta' = 0$ or $1$), forward rapidity of the photon ($\eta^\gamma \gtrsim 1$), smaller outgoing electron transverse momenta ($p_T^\prime \lesssim 2\,{\rm GeV}$ at electron midrapidity and $p_T^\prime \lesssim 4\,{\rm GeV}$ at electron forward rapidity), and larger photon transverse momenta ($p_T^\gamma \gtrsim 5\,{\rm GeV}$, although at electron forward rapidity, $p_T^\gamma \gtrsim 1\,{\rm GeV}$ can still give sizable asymmetries).  If the CM energy is further increased to $\sqrt{s}=141\,{\rm GeV}$, we found the asymmetry is zero everywhere except at the most forward kinematics of $\eta'=1$ and $\eta^\gamma \gtrsim 2$.

Finally, we would like to mention that there exists, at least in principle, the possibility to experimentally disentangle the two distinct channels, Compton (C) and Interference (I), in Eq.~\eqref{e:AUT} using an electron as well as a positron beam. We explicitly checked that both the Compton and Interference channels make similar-sized contributions to $A_{UT}^{\gamma {\rm SIDIS}}$ in Figs.~\ref{f:AUTscen0}, \ref{f:AUTscen1}, \ref{f:AUTscen1_63}, and neither of the channels dominates the other over the full range of the kinematics. Specifically, a \emph{beam charge asymmetry} will cleanly probe the Interference channel on its own. This would be important since the Interference channel is generated by \emph{valence-type} $q-\bar{q}$ (instead of $q+\bar{q}$) combinations of multi-parton correlation functions (see Eq.~\eqref{e:BHCI}). Such information on $q-\bar{q}$ distributions would be most helpful for separating quark- and antiquark correlation functions. We note that there is the opportunity for a positron beam at Jefferson Lab~\cite{Afanasev:2019xmr,Arrington:2021alx} and the EIC~\cite{AbdulKhalek:2021gbh}, 
and we consider the beam charge asymmetry to be an interesting physics measurement at those facilities.

\vspace{-0.15cm}
\section{Conclusions} \label{s:concl}
We have numerically analyzed the transverse single-spin asymmetry $A_{UT}^{\gamma {\rm SIDIS}}$ in the semi-inclusive deep-inelastic production of isolated photons from electron-proton scattering.  We utilized realistic models for the quark-gluon-quark correlation functions $F_{FT}(x,x')$ and $G_{FT}(x,x')$ based on the Sivers function and a constraint from the $d_2$ matrix element computed by lattice QCD.  We found that $\big |A_{UT}^{\gamma {\rm SIDIS}}\big |$ is the largest at smaller center-of-mass energies (e.g., the $\sqrt{s}=29\,{\rm GeV}$ configuration at the EIC) with the outgoing electron at mid or backward rapidity, photon at mid to forward rapidity, smaller electron transverse momentum, higher photon transverse momentum, and with the electron and photon produced in the same direction (namely, azimuthal angles of $0$ or $\pi$). A user-friendly google colab notebook is available~\cite{AUTgamSIDIS_lib} for the reader to explore other choices for model parameters for $F_{FT}(x,x')$ and $G_{FT}(x,x')$ as well as alternative experimental configurations. Our results provide encouraging evidence that $A_{UT}^{\gamma {\rm SIDIS}}$ will be measurable at the EIC, which would provide unprecedented information on $F_{FT}(x,x')$ and $G_{FT}(x,x')$ for their full support in $x,x'$.  

\section*{Acknowledgments}  
\noindent This work was supported by the National Science Foundation under Grants  No.~PHY-2308567 (D.P., M.H., J.M., J.P.), and No.~PHY-2310031, No.~PHY-2335114 (A.P.), the U.S. Department of Energy contract No.~DE-AC05-06OR23177, under which Jefferson Science Associates, LLC operates Jefferson Lab (A.P.), and by Deutsche Forschungsgemeinschaft (DFG) through the Research Unit FOR 2926 (Project No. 409651613) (D.R.~and M.S.).

\newpage

\appendix

\section{Partonic Cross Sections for the Transversely Polarized Process}\label{s:partCS}
In this appendix we collect the explicit formulas for the partonic cross sections that appear in the final result for the transversely polarized process given in Eq.~\eqref{e:sigmaUT}. We separate them by the type of pole contribution, i.e., hard pole~(HP) and soft-fermion pole (SFP), by the kind of linear combination of the quark-gluon-quark functions ($\pm$), by the type of azimuthal spin correlation ($\phi^\prime$ or $\phi^\gamma$), and by the production channel (Compton or Interference). In total there are 16 different hard scattering coefficients which we list below. We remind the reader that these results were already obtained in Ref.~\cite{Albaltan:2019cyc}. We express our results in the same dimensionless kinematic variables that were used in this reference and first introduced in Ref.~\cite{Brodsky:1972}. The variables can be defined via the relations
\begin{eqnarray}
    P\cdot l\,=\,\tfrac{1}{2}Q^2\alpha\,,&&\,P\cdot l^\prime\,=\,\tfrac{1}{2}Q^2\alpha^\prime\,,\nonumber\\
    l\cdot P_\gamma\,=\,\tfrac{1}{2}Q^2\beta\,,&&\,l^\prime\cdot P_\gamma\,=\,\tfrac{1}{2}Q^2\beta^\prime\,,\nonumber\\
    P\cdot P_\gamma\,=\,\tfrac{1}{2}Q^2\gamma\,,&&\,\tilde{Q}^2\,=\,\left(1-\beta+\beta^\prime\right)Q^2\,.\label{e:BrodskyVars}
\end{eqnarray}
Note that these variables are also related to $x_B,\,\tilde{x}_B$, which were defined in the beginning of Sec.~\ref{s:theory}, as follows
\begin{equation}
    x_B=\frac{1}{\alpha-\alpha^\prime-\gamma}\,,\quad\quad
    \tilde{x}_B=\frac{1-\beta+\beta^\prime}{\alpha-\alpha^\prime}\,.\label{e:Bjorken2Brodsky}
\end{equation}
We use these relations to eliminate $\gamma,\,\tilde{x}_B$ in favor of $\alpha,\,\alpha^\prime,\,\beta,\,\beta^\prime,\,x_B$ in our results for the partonic cross sections below. Using these kinematic variables, we can also write explicitly the azimuthal spin correlations in a compact form
\begin{equation}
    \epsilon^{Pll^\prime S}=\frac{Q^3}{2}\sqrt{\alpha \alpha^\prime\left(1-\beta+\beta^\prime\right)}\,\sin\left(\phi_S-\phi^\prime\right), \quad\quad
    \epsilon^{PlP_\gamma S}=\frac{Q^3}{2}\sqrt{\alpha\beta\gamma}\,\sin\left(\phi_S-\phi^\gamma\right)\,.\label{e:azimuthal}
\end{equation}
Moreover, we introduce several short-hand notations for the different denominators that will appear in our formulas
\begin{equation}
    D_1=1-\beta+\beta^\prime\,,\quad\quad
    D_2=1-x_B\left(\alpha-\alpha^\prime\right), \quad\quad
    D_3=1-\beta+\beta^\prime-x_B\left(\alpha-\alpha^\prime\right)\,.\label{e:denominators}
\end{equation}
Making use of the two simplifications described in the first full paragraph after Eq.\eqref{e:BHCI} gives us the following hard factors.

\paragraph{Hard Pole Contributions}
The hard pole contributions for the ``$+$" combination of $F_{FT}$ and $G_{FT}$ are given by the following partonic cross sections
\begin{eqnarray}
    \hat{\sigma}_{\rm HP,+}^{C,\phi^\prime}&\!\!=\!\!&-\frac{16 \left(\alpha ^2+\left(\alpha '\right)^2\right) x_B^2}{\alpha  D_1 D_3^2}\,,\nonumber\\
    \hat{\sigma}_{\rm HP,+}^{I,\phi^\prime}&\!\!=\!\!&\frac{16 \left(\alpha ^2+\left(\alpha '\right)^2\right) x_B^2 \left(D_1 \alpha ' \beta ' x_B+\alpha  \left(\beta ^2-(\beta +1) \beta '\right)
   x_B+D_1 \beta '\right)}{\alpha  \beta  \beta ' D_1 D_3^2}\,,\nonumber\\
   \hat{\sigma}_{\rm HP,+}^{C,\phi^\gamma}&\!\!=\!\!&\frac{16 \left(\alpha -\alpha '\right) \left(\alpha ^2+\left(\alpha '\right)^2\right) x_B^3}{\alpha  D_1 D_2 D_3^2}\,,\nonumber\\
   \hat{\sigma}_{\rm HP,+}^{I,\phi^\gamma}&\!\!=\!\!&-\frac{16 \left(\alpha ^2+\left(\alpha '\right)^2\right) x_B^3 \left(\alpha  \beta -\alpha ' \beta '\right)}{\alpha  \beta \beta ' D_3^2 }\,.\label{e:HPplus}
\end{eqnarray}
And for the ``$-$" combination of $F_{FT}$ and $G_{FT}$ we find the following, slightly more lengthy, partonic cross sections
\begin{eqnarray}
    \hat{\sigma}_{\rm HP,-}^{C,\phi^\prime}&\!\!=\!\!&-\frac{16 \left(2 D_1 \alpha ' \left(\beta '+1\right) x_B+2 x_B^2 \left(\alpha ^2 (\beta -1) \beta -\alpha \alpha ' \beta   \left(2 \beta
   '+1\right)+\left(\alpha '\right)^2 \left(\beta '+1\right)^2\right)+D_1^2\right)}{\alpha  D_1^2 \left(1-D_2\right) D_3^2}\,,\nonumber\\
   \hat{\sigma}_{\rm HP,-}^{I,\phi^\prime}&\!\!=\!\!&\frac{8}{\alpha  \beta\beta ' D_1 \left(1-D_2\right) D_3^2 }\left(D_1 x_B \left(6 \alpha ' \left(\beta '\right)^2+2 \alpha  \beta ^2-\beta ' \left(2 \beta  \left(\alpha '+\alpha \right)-6 \alpha'+\alpha  \left(\alpha  \alpha ' x_B^2+2\right)\right)\right)\nonumber\right.\\
   &&\left.\,+x_B^2 \left(-4 \alpha ^3 (\beta -1) \beta ^2 x_B-4 \alpha  \alpha ' \beta '\left(2 \beta  \beta '+\beta '+\beta +\alpha ' \left(3 \beta  \beta '+\beta '+2 \beta +1\right) x_B+1\right)\right.\right.\nonumber\\
   &&\left.\left.\,+\,4 \left(\alpha '\right)^2\beta ' \left(\beta '+1\right) \left(2 \beta '-\beta +\,\alpha ' \left(\beta '+1\right) x_B+2\right)\right.\right.\nonumber\\
   &&\left.\left.\,+\,\alpha ^2 \left(4 (\beta -1) \beta^2+\alpha ' \left(\beta '\right)^2 x_B+(\beta  (12 \beta -1)+1) \alpha ' \beta ' x_B\right)\right)+2 D_1^2 \beta ' \right),\nonumber\\
   \hat{\sigma}_{\rm HP,-}^{C,\phi^\gamma}&\!\!=\!\!&-\frac{16 x_B }{\alpha  D_1^2 D_2 \left(1-D_2\right) D_3^2}\left(2 D_1 \left(\alpha '\right)^2 \left(\beta '+1\right) x_B\nonumber\right.\\
   &&\left.\,+\,2 x_B \left(-\alpha ^3 (\beta -1)^2 x_B-\alpha \left(\alpha '\right)^2 \left(\left(\beta '\right)^2+2 \beta  \beta '+\beta '+\beta \right) x_B+\left(\alpha '\right)^3 \left(\beta'+1\right)^2 x_B\nonumber\right.\right.\\
   &&\left.\left.\,+\,\alpha ^2 \left(\beta '\left(1-\beta\right)+(2 \beta -1) \alpha ' \beta ' x_B+(\beta -1) \left(\beta +\beta  \alpha 'x_B-1\right)\right)\right)+D_1^2 \left(\alpha '-\alpha \right)\right),\nonumber\\
   \hat{\sigma}_{\rm HP,-}^{I,\phi^\gamma}&\!\!=\!\!&\frac{16 x_B }{\alpha  \beta \beta ' D_1 \left(1-D_2\right) D_3^2 }\left(\beta ' \left(-\beta  \left(\alpha '+\alpha \right) \left(x_B \left(\alpha '-\alpha +4 \alpha  \alpha '
   x_B\right)-D_1+D_3+2\right)\nonumber\right.\right.\\
   &&\left.\left.\,+\,\beta ^2 \left(\alpha '+2 \alpha  \left(\alpha  x_B \left(3 \alpha ' x_B-1\right)+1\right)\right)+\alpha '
   \left(2 \alpha ' x_B \left(\alpha ' x_B+1\right)+1\right)\right)\nonumber\right.\\
   &&\left.\,+\,\alpha ' \left(\beta '\right)^3 \left(2 \alpha ' x_B \left(\alpha '
   x_B+1\right)+1\right)+\left(\beta '\right)^2 \left(2 \alpha ' \left(-\beta +\alpha ' x_B \left(-\beta +2 \alpha ' x_B-3 \alpha  \beta 
   x_B+2\right)+1\right)-\alpha  \beta \right)\nonumber\right.\\
   &&\left.-\,\alpha  (\beta -1)^2 \beta  \left(2 \alpha  x_B \left(\alpha 
   x_B-1\right)+1\right)\right)\,.\label{e:HPminus}
\end{eqnarray}

\paragraph{Soft-Fermion Pole Contributions}
The soft fermion pole contributions for the ``$+$" combination of $F_{FT}$ and $G_{FT}$ are given by the following partonic cross sections
\begin{eqnarray}
    \hat{\sigma}_{\rm SFP,+}^{C,\phi^\prime}&\!\!=\!\!&\frac{16 \left(\beta ' \left(\beta '+2\right)+(\beta -2) \beta -2 x_B \left((\beta -1) \alpha '+\alpha  \left(\beta
   '+1\right)\right)+\left(\alpha ^2+\left(\alpha '\right)^2\right) x_B^2+2\right)}{\alpha  D_1 D_2 D_3}\,,\nonumber\\
   \hat{\sigma}_{\rm SFP,+}^{I,\phi^\prime}&\!\!=\!\!&-\frac{16 \left(\beta ' \left(\beta '+2\right)+(\beta -2) \beta -2 x_B \left((\beta -1) \alpha '+\alpha  \left(\beta
   '+1\right)\right)+\left(\alpha ^2+\left(\alpha '\right)^2\right) x_B^2+2\right)}{\alpha   \beta 'D_1 D_2}\,,\nonumber\\
   \hat{\sigma}_{\rm SFP,+}^{C,\phi^\gamma}&\!\!=\!\!&-\frac{16 \left(\alpha -\alpha '\right) x_B \left(\beta ' \left(\beta '+2\right)+(\beta -2) \beta -2 x_B \left((\beta -1) \alpha '+\alpha 
   \left(\beta '+1\right)\right)+\left(\alpha ^2+\left(\alpha '\right)^2\right) x_B^2+2\right)}{\alpha  D_1 D_2^2 D_3}\,,\nonumber\\
   \hat{\sigma}_{\rm SFP,+}^{I,\phi^\gamma}&\!\!=\!\!&\frac{16 x_B \left(\alpha  \beta '-\beta  \alpha '\right) \left(\beta ' \left(\beta '+2\right)+(\beta -2) \beta -2 x_B \left((\beta -1) \alpha
   '+\alpha  \left(\beta '+1\right)\right)+\left(\alpha ^2+\left(\alpha '\right)^2\right) x_B^2+2\right)}{\alpha  \beta   \beta 'D_1 D_2^2}\,.\label{e:SFPplus}
\end{eqnarray}
And for the ``$-$" combination of $F_{FT}$ and $G_{FT}$ we find the following,
\begin{eqnarray}
    \hat{\sigma}_{\rm SFP,-}^{C,\phi^\prime}&\!\!=\!\!&\frac{16 \left(x_B \left(\alpha ' \left(\beta ^2+\left(\beta '+1\right)^2\right)-\alpha  \left(\left(\beta '+1\right)^2+(\beta -2) \beta
   \right)\right)+D_1 \left(\beta ^2+\left(\beta '+1\right)^2\right)\right)}{\alpha  D_1^2 D_2 D_3}\,,\nonumber\\
   \hat{\sigma}_{\rm SFP,-}^{I,\phi^\prime}&\!\!=\!\!&-\frac{16 \left(\alpha  D_1 \left(\beta ^2+\left(\beta '\right)^2\right) x_B+D_2 \beta ' \left(\beta ^2+\left(\beta
   '+1\right)^2\right)\right)}{\alpha  \beta   \beta 'D_1 D_2}\,,\nonumber\\
   \hat{\sigma}_{\rm SFP,-}^{C,\phi^\gamma}&\!\!=\!\!&\frac{16 x_B }{\alpha  D_1^2 D_2^2 D_3}\left(\alpha ^2 \left(\left(\beta '\right)^2+(\beta -1)^2\right) x_B-2 \alpha  \alpha ' \left(\left(\beta '+1\right)^2+(\beta -2)
   \beta \right) x_B\nonumber\right.\\
   &&\left.+\left(\alpha '\right)^2 \left(\beta ^2+\left(\beta '+1\right)^2\right) x_B+D_1 \alpha ' \left(\beta ^2+\left(\beta
   '+1\right)^2\right)-\alpha  D_1 \left(\left(\beta '\right)^2+(\beta -1)^2\right)\right),\nonumber\\
   \hat{\sigma}_{\rm SFP,-}^{I,\phi^\gamma}&\!\!=\!\!&\frac{16 x_B \left(\alpha  \beta  \left(\left(\beta '\right)^2+(\beta -1)^2\right)-\alpha ' \beta ' \left(\beta ^2+\left(\beta
   '+1\right)^2\right)\right)}{\alpha  \beta   \beta 'D_1 D_2}\,.\label{e:SFPminus}
\end{eqnarray}

\end{document}